\documentclass[showpacs,showkeys,nofootinbib]{revtex4}%
\usepackage{graphicx}
\usepackage{epsfig}
\usepackage{bm}
\usepackage{amsmath}
\usepackage{amsfonts}
\usepackage{amssymb}
\usepackage{subfigure}%
\begin{document}
\title{Pion pole light-by-light contribution to $g-2$ of the muon in a nonlocal
chiral quark model}
\author{Alexander E. Dorokhov}
\email{dorokhov@theor.jinr.ru}
\affiliation{Joint Institute for Nuclear Research, Bogoliubov Laboratory of Theoretical
Physics, 141980, Moscow region, Dubna, Russia}
\author{Wojciech Broniowski}
\email{Wojciech.Broniowski@ifj.edu.pl}
\affiliation{Henryk Niewodnicza\'{n}ski Institute of Nuclear Physics, Polish Academy of Sciences,
PL-31342~Krak\'{o}w,
Poland; Institute of Physics, Jan Kochanowski University, PL-25406~Kielce, Poland}

\begin{abstract}
We calculate the pion pole term of the light-by-light contribution to the $g-2$ of the
muon in the framework of an effective chiral quark model with instanton-like nonlocal
quark--quark interaction. The full kinematic dependence of the pion-photon transition
form factors is taken into account. The dependence of form factors on the pion virtuality
decreases the result by about 15\% in comparison to the calculation where this dependence is
neglected. Further, it is demonstrated that the QCD constraints suggested by Melnikov and
Vainshtein are satisfied within the model. The corresponding contributions originate from
the box diagram as well from the pion-pole term. Our chiral nonlocal model result for the
pion-pole light-by-light contribution to $(g-2)/2$ of the muon is $\left(  6.3-6.7\right)
\cdot10^{-10}$, which is in the ball park of other effective-model calculations.

\end{abstract}
\maketitle

\section{Introduction}

\label{1} \setcounter{equation}{0}

The E821 experiment at the Brookhaven National Laboratory has recently
measured the anomalous magnetic moment of the muon, $a_{\mu}\equiv(g_{\mu
}-2)/2$, with the final result \cite{Bennett:2006fi}:
\begin{equation}
a_{\mu}^{\mathrm{exp}}=11\,659\,208.0(6.3)\cdot10^{-10}.
\end{equation}
This unprecedented accuracy, with yet better precision expected in the planned
experiments at BNL, JPARC, and FNAL \cite{Hertzog:2007hz}, maintains the live
interest in obtaining more accurate theoretical predictions for $a_{\mu}$
within the standard model, for reviews see, e.g.,
\cite{Miller:2007kk,Passera:2007fk,Dorokhov:2005ff,Jegerlehner:2007xe}. The
challenge is to obtain the theoretical uncertainty lower than the
uncertainties for the nearest-future experiments, which will supply a powerful
test for possible effects of contributions from the New Physics.

\begin{figure}[th]
\includegraphics[width=10cm]{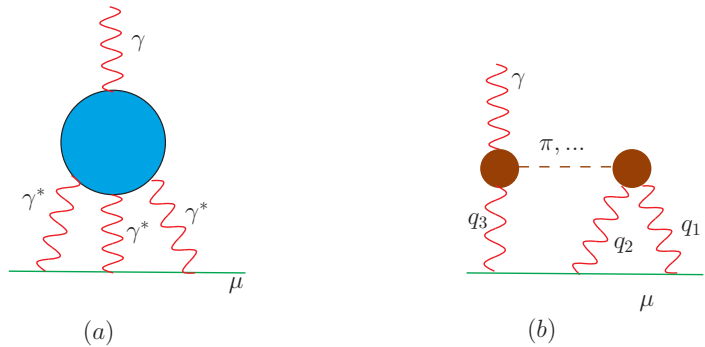}\caption{(Color online) (a) Hadronic
light-by-light contribution to $a_{\mu}$. The bottom line is the muon,
the wavy lines are the photons, and the circle depicts the hadronic part. (b)
The meson pole contribution to $a_{\mu}$. The circles represent the virtual
meson to $\gamma^{\ast}\gamma^{\ast}$ transition form factors.}%
\label{LbLfig}%
\end{figure}

In the standard model the electromagnetic (QED), electroweak (EW), and hadronic (Had) effects
contribute to $a_{\mu}$. Taking into account the recent progress with the QED
calculations and the latest result for $a_{e}$ one obtains
\cite{Passera:2006gc}
\begin{equation}
a_{\mu}^{\mathrm{QED}}=11\,658\,471.809(0.016)\cdot10^{-10}\,.\label{QED}%
\end{equation}
The electroweak contribution to $a_{\mu}$ is also known
accurately~\cite{Czarnecki:2002nt},
\begin{equation}
a_{\mu}^{\mathrm{EW}}\mathrm{=15.4(0.2)\cdot10^{-10}.}\label{EW}%
\end{equation}
The main source of the theoretical uncertainty is the hadronic contribution.
There are three types of the leading hadronic contributions: the vacuum
polarization, its next-to-leading
folding with the QED and EW sectors, and the light-by-light (LbL) scattering
process (Fig.~\ref{LbLfig}). A recent phenomenological reanalysis of the
contribution of the full hadron vacuum polarization insertion into the
electromagnetic vertex of the muon \cite{Eidelman:2007zz} gives
\begin{equation}
a_{\mu}^{\mathrm{Had,LO}}=690.9(4.4)\cdot10^{-10}.\label{HLO}%
\end{equation}
The most recent estimate of the higher-order (HO) hadronic contributions performed
in \cite{Hagiwara:2003da} provides the result
\begin{equation}
a_{\mu}^{\mathrm{Had,HO}}\mathrm{=-9.8(0.1)\cdot10^{-10}.}\label{HNLO}%
\end{equation}
For the LbL contributions there are several model-dependent estimates:
\begin{eqnarray}
&a_{\mu}^{\mathrm{{Had,LbL}}} &  =8.3(3.2)\cdot10^{-10}%
\ \ \ \ \ \ [11,12],\label{lblB}\\
&a_{\mu}^{\mathrm{{Had,LbL}}} &  =8.9(1.7)\cdot10^{-10}\ \ \ \ \ \  [13]
,\label{lblH}\\
&a_{\mu}^{\mathrm{{Had,LbL}}} &  =13.6(2.5)\cdot10^{-10}\ \ \ \ \  [14]
.\label{lblM}%
\end{eqnarray}

It is clear that the leading-order (LO) hadronic contribution (\ref{HLO}) dominates
in the absolute value. However, the theoretical error introduced by the LbL
process (\ref{lblB})-(\ref{lblM}) is of the same order as that for the leading
term. Moreover, the error in (\ref{HLO}) is phenomenologically well controlled
and may be improved by a factor of 2 or so if more experimental data on
$e^{+}e^{-}\left(  \tau\right)  \rightarrow\mathit{hadrons}$ should appear
\cite{Eidelman:2007zz}. The precision for $a_{\mu}^{\mathrm{Had,HO}}$ is quite
enough for the nearest-future experiments. Unfortunately, the hadronic LbL
contribution cannot be related to any other observable and hence we must rely on
purely theoretical model framework in order to estimate it. Predictions based
on chiral models (\ref{lblB}), (\ref{lblH}) (also in \cite{Knecht:2001qf}) are
compatible with one another and are much lower than the results obtained in
(\ref{lblM}) (also in \cite{Bartos:2001pg,Pivovarov:2001mw}). Furthermore, the
errors given in (\ref{lblB})-(\ref{lblM}) are difficult to estimate and it is
not easy to improve their quality. These errors should include the model
dependence, the dependence on model parameters, and the quality of the model
assumptions. The last point, difficult to assess, is most important for our
further discussion presented in this paper.

We digress that at present the calculation of the LbL contribution within
lattice QCD is a quite difficult task \cite{Hayakawa:2005eq}, hence we cannot
use lattices yet as a reliable source of information for the considered problem.

The present work is devoted to the calculation of the pion-pole contribution
of the hadronic light-by-light scattering to $a_{\mu}$ within the nonlocal
chiral quark model (N$\chi$QM) \cite{Anikin:2000rq,Dorokhov:2003kf} motivated
by the instanton model of the QCD vacuum
\cite{Callan:1977gz,Shuryak:1981ff,Diakonov:1985eg}. In \cite{Dorokhov:2003kf}
the vector and axial-vector correlators were calculated and then these results
were applied to compute the hadronic vacuum polarization contribution $a_{\mu
}^{\mathrm{{Had,LO}}}$ \cite{Dorokhov:2004ze}. Later on the three-point $PVV$
and $VVA$ correlators were analyzed in
\cite{Dorokhov:2002iu,Dorokhov:2005pg,Dorokhov:2005hw} and the contribution of
the $\gamma\gamma^{\ast}Z^{\ast}$ vertex to $a_{\mu}$ was estimated in
\cite{Dorokhov:2005ff}. The present calculation of the LbL contribution
exhibits a few important improvements compared to the previous calculations in
effective quark models. First of all, in the pion-pole contribution, which
dominates the hadronic LbL part, we take into account the full kinematic
dependence of the pion-photon form factors, including their dependence on the
pion virtuality. The inclusion of this dependence diminishes the results for
the contribution to $a_{\mu}$ by about 15\% compared to the calculation with
no dependence on the pion virtuality. Importantly, our approach is consistent
with the low-energy theorems and with the QCD constraints. In particular, we
demonstrate how the QCD constraint considered by Melnikov and Vainshtein
\cite{Melnikov:2003xd} is satisfied at high photon momenta within N$\chi$QM,
with the crucial role of the box diagram. We also show that in general the
pion-pole contribution exhibits no enhancement as it was assumed in the model
\cite{Melnikov:2003xd} for the pion exchange in the LbL amplitude.

The next Section contains the definitions of the LbL amplitude, its pion-pole
contribution, and the QCD constraint for the amplitude discussed in
\cite{Melnikov:2003xd}. Sections III and IV describe the nonlocal chiral quark
model based on instanton-like dynamics (N$\chi$QM), introduces conserved
vector and axial-vector currents, as well as the relevant $PVV$ amplitude.
This part, developed in previous works, is included for the completeness of
the paper. Numerical results for the pion pole contribution to LbL and the
comparison to other calculations are shown in Sections V and VI. In Section
VII the Melnikov-Vainshtein constraint is proven within N$\chi$QM. Conclusions
are given in the last Section.

\section{Light-by-Light amplitude\label{I}}

The amplitude for the light-by-light scattering is defined as\footnote{In the
following definitions we follow closely the notation used in
\cite{Melnikov:2003xd}.}
\begin{eqnarray}
&&  \mathcal{M}=\alpha^{2}N_{c}\,\mathrm{Tr}\,[{\hat{Q}}^{4}]\,\mathcal{A}%
=\alpha^{2}N_{c}\,\mathrm{Tr}\,[{\hat{Q}}^{4}]\,\mathcal{A}_{\mu_{1}\mu_{2}%
\mu_{3}\gamma\delta}\epsilon_{1}^{\mu_{1}}\epsilon_{2}^{\mu_{2}}\epsilon
_{3}^{\mu_{3}}f^{\gamma\delta}\nonumber\\
&&  =-e^{3}\!\int\!\mathrm{d}^{4}x\,\mathrm{d}^{4}y\,\mathrm{e}^{-iq_{1}%
x-iq_{2}y}\,\epsilon_{1}^{\mu_{1}}\epsilon_{2}^{\mu_{2}}\epsilon_{3}^{\mu_{3}%
}<0|T\left\{  j_{\mu_{1}}(x)\,j_{\mu_{2}}(y)\,j_{\mu_{3}}(0)\right\}
|{\gamma}>,\label{eqcala}%
\end{eqnarray}
where $q_{i}$ and $\epsilon_{i}$ are the momenta and the polarization vectors
of photons, $j_{\mu}$ is the hadronic electromagnetic current defined
explicitly below within the nonlocal chiral quark model, and $\hat{Q}$ is the
quark charge operator. The photon momenta are taken to be incoming, with $\sum
q_{i}=0$. One of the photons represents the external magnetic field and can be
regarded as a real photon with a vanishingly small momentum, $q_{4}$. Due to
the gauge invariance the light-by-light scattering amplitude is proportional
to the field strength tensor of the soft photon, $f^{\gamma\delta}%
\!=\!q_{4}^{\gamma}\epsilon_{4}^{\delta}-q_{4}^{\delta}\epsilon_{4}^{\gamma}$.
Neglecting the quadratic and higher powers of $q_{4}$, the tensor amplitude
$\mathcal{A}_{\mu_{1}\mu_{2}\mu_{3}\gamma\delta}$ may be considered as a
function of the photon virtualities $q_{i}^{2}$, $i=1,2,3$ under the condition
$q_{1}+q_{2}+q_{3}=0$.

In general the LbL amplitude is a rather complicated object to analyze.
However, it is possible to generate different hierarchies for the components
of the amplitude. The dominant contribution comes from the pole in the
pseudoscalar channel, which is enhanced by the very small value of the pion mass (Fig.
\ref{LbLfig}b). It is also leading in terms of the large-$N_{c}$ counting,
where $N_{c}$ is the number of colors. Moreover, the leading-$N_{c}$
contributions from other mesonic channels are suppressed by much larger meson masses.

It can be shown, see \cite{Aldins:1970id,Bijnens:1995xf,Knecht:2001qf}, that
the leading contribution from the neutral pseudoscalar meson exchange to
$a_{\mu}$ is given by
\begin{eqnarray}
&a_{\mu}^{\mbox{\tiny{LbL;$\pi^0$}}} &  =-e^{6}\int{\frac{d^{4}q_{1}}%
{(2\pi)^{4}}}\int{\frac{d^{4}q_{2}}{(2\pi)^{4}}}\,\frac{1}{q_{1}^{2}q_{2}%
^{2}q_{3}^{2}[(p+q_{1})^{2}-m^{2}][(p-q_{2})^{2}-m^{2}]}\nonumber\\
&&  \quad\quad\times\left[  \mathcal{F}_{\pi_{0}^{\ast}\gamma^{\ast}%
\gamma^{\ast}}(q_{2}^{2};q_{1}^{2},q_{3}^{2}){\frac{G_{P}\ }{g_{\pi q}%
^{2}\left(  1-G_{P}J_{pp}\left(  q_{2}^{2}\right)  \right)  }}\mathcal{F}%
_{\pi_{0}^{\ast}\gamma^{\ast}\gamma^{\ast}}(q_{2}^{2};q_{2}^{2},0)\ T_{1}%
(q_{1},q_{2};p)\right.  \nonumber\\
&&  \quad\quad\quad+\left.  \mathcal{F}_{\pi_{0}^{\ast}\gamma^{\ast}%
\gamma^{\ast}}(q_{3}^{2};q_{1}^{2},q_{2}^{2}){\frac{G_{P}\ }{g_{\pi q}%
^{2}\left(  1-G_{P}J_{pp}\left(  q_{3}^{2}\right)  \right)  }}\ \mathcal{F}%
_{\pi_{0}^{\ast}\gamma^{\ast}\gamma^{\ast}}(q_{3}^{2};(q_{1}+q_{2}%
)^{2},0)T_{2}(q_{1},q_{2};p)\right]  \,,\label{PionPole}%
\end{eqnarray}
where $m$ denotes the muon mass ($p^{2}=m^{2}$) and the kinematic factors are
\cite{Knecht:2001qf}
\begin{eqnarray}
&T_{1}(q_{1},q_{2};p) &  ={\frac{16}{3}}\,(p\cdot q_{1})\,(p\cdot
q_{2})\,(q_{1}\cdot q_{2})\,-\,{\frac{16}{3}}\,(p\cdot q_{2})^{2}\,q_{1}%
^{2}\nonumber\\
&&  \!\!\!\!\!-\,{\frac{8}{3}}\,(p\cdot q_{1})\,(q_{1}\cdot q_{2})\,q_{2}%
^{2}\,+\,8(p\cdot q_{2})\,q_{1}^{2}\,q_{2}^{2}\,-\,{\frac{16}{3}}(p\cdot
q_{2})\,(q_{1}\cdot q_{2})^{2}\nonumber\\
&&  \!\!\!\!\!+\,{\frac{16}{3}}\,m^{2}\,q_{1}^{2}\,q_{2}^{2}\,-\,{\frac{16}{3}%
}\,m^{2}\,(q_{1}\cdot q_{2})^{2}\,,\label{t1}\\
&T_{2}(q_{1},q_{2};p) &  ={\frac{16}{3}}\,(p\cdot q_{1})\,(p\cdot
q_{2})\,(q_{1}\cdot q_{2})\,-\,{\frac{16}{3}}\,(p\cdot q_{1})^{2}\,q_{2}%
^{2}\nonumber\\
&&  \!\!\!\!\!+\,{\frac{8}{3}}\,(p\cdot q_{1})\,(q_{1}\cdot q_{2})\,q_{2}%
^{2}\,+\,{\frac{8}{3}}\,(p\cdot q_{1})\,q_{1}^{2}\,q_{2}^{2}\,\nonumber\\
&&  \!\!\!\!\!+\,{\frac{8}{3}}\,m^{2}\,q_{1}^{2}\,q_{2}^{2}\,-\,{\frac{8}{3}%
}\,m^{2}\,(q_{1}\cdot q_{2})^{2}\,.\label{t2}%
\end{eqnarray}
The $PVV$ amplitude $\mathcal{F}_{\pi_{0}^{\ast}\gamma^{\ast}\gamma^{\ast}%
}(q_{3}^{2};q_{1}^{2},q_{2}^{2})$ for the virtual pion and photons, as well as
the meson propagator in the pseudoscalar channel, $\left(  1-G_{P}J_{pp}\left(
q_{2}^{2}\right)  \right)  ^{-1}$, will be discussed in Sect.~\ref{IV}.

\begin{figure}[th]
\includegraphics[width=10cm]{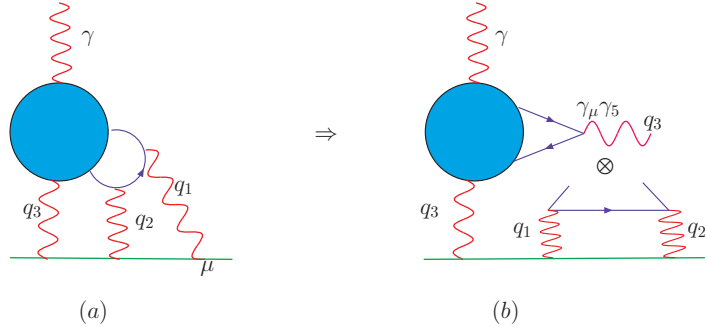}\caption{(Color online) In the limit
$q_{1}^{2}\approx q_{2}^{2}\gg q_{3}^{2}$ considered in \cite{Melnikov:2003xd}
the LbL amplitude (a) factorizes in the leading twist into the $VVA$ soft
hadronic part and hard coefficient function, as indicated in (b). }%
\label{MVfig}%
\end{figure}

Recently, an important constraint was introduced by Melnikov and Vainshtein
(MV) \cite{Melnikov:2003xd}, who argued with the help of the operator product
expansion (OPE) that in the specific kinematic limit, $q_{1}^{2}\approx
q_{2}^{2}\equiv q^{2}\gg q_{3}^{2}$, the amplitude has the asymptotic form
(see Fig.~\ref{MVfig})
\begin{equation}
\mathcal{A}_{\mu_{1}\mu_{2}\mu_{3}\gamma\delta}f^{\gamma\delta}=\frac{8}%
{{\hat{q}}^{2}}\epsilon_{\mu_{1}\mu_{2}\delta\rho}{\hat{q}}^{\delta}%
\!\!\sum_{a=3,8,0}\!\!W^{(a)}\left\{  w_{L}^{(a)}\!(q_{3}^{2})\,q_{3}^{\rho
}q_{3}^{\sigma}\tilde{f}_{\sigma\mu_{3}}+w_{T}^{(a)}\!(q_{3}^{2})\!\left(
-q_{3}^{2}\tilde{f}_{\mu_{3}}^{\rho}\!+\!q_{3\mu_{3}}q_{3}^{\sigma}\tilde
{f}_{\sigma}^{\rho}\!-\!q_{3}^{\rho}q_{3}^{\sigma}\tilde{f}_{\sigma\mu_{3}%
}\right)  \right\}  +\cdots,\label{MVlimit}%
\end{equation}
where $\tilde{f}_{\alpha\beta}=\frac{1}{2}\varepsilon_{\alpha\beta\gamma
\delta}f^{\gamma\delta},$ ${\hat{q}=}\left(  q_{1}-q_{2}\right)  /2$ and no
hierarchy between $q_{3}^{2}$ and $\Lambda_{\mathrm{QCD}}^{2}$ is assumed. The
weights $W^{(a)}$ are defined as
\begin{eqnarray}
&&  W^{(a)}=\frac{\left(  \mathrm{Tr}\,[\lambda_{a}\hat{Q}^{2}]\right)  ^{2}%
}{\mathrm{Tr}\,[\lambda_{a}^{2}]\mathrm{Tr}\,[\hat{Q}^{4}]}\,;\label{weights}%
\\[1mm]
&&  W^{(3)}=\frac{1}{4}\,,\quad W^{(8)}=\frac{1}{12}\,,\quad W^{(0)}=\frac
{2}{3}\,.\nonumber
\end{eqnarray}
The invariant functions $w_{L}^{(a)}\!(q^{2})$ and $w_{T}^{(a)}\!(q^{2})$ are
related to the triangle amplitude $T_{\gamma\rho}^{(a)}$ that involves the
axial current $j_{5\rho}^{(a)}$ and two electromagnetic currents, one with
momentum $q$ and the other one (the external magnetic field) with the
vanishing momentum (see the relevant part of Fig. \ref{MVfig}b),
\begin{equation}
T_{\mu\rho}^{(a)}\!=\!\!=i\,\langle0|\!\int\!\mathrm{d}^{4}z\,\mathrm{e}%
^{iqz}T\{j_{5\rho}^{(a)}(z)\,j_{\mu}(0)\}|\gamma\rangle.\label{Triangle}%
\end{equation}
It is shown in \cite{Vainshtein:2002nv} that $T_{\gamma\rho}^{(a)}$ can be
written as
\begin{eqnarray}
&&  T_{\mu\rho}^{(a)}=-\frac{ie\,N_{c}\mathrm{Tr}\,[\lambda_{a}\hat{Q}^{2}%
]}{4\pi^{2}}\left\{  w_{L}^{(a)}\!(q^{2})\,q_{\rho}q^{\sigma}\tilde{f}%
_{\sigma\mu}+\right.  \nonumber\\
&&  \left.  +w_{T}^{(a)}\!(q^{2})\!\left(  -q^{2}\tilde{f}_{\mu\rho}%
\!+\!q_{\mu}q^{\sigma}\tilde{f}_{\sigma\rho}\!-\!q_{\sigma}q^{\sigma}\tilde
{f}_{\sigma\mu}\right)  \right\}  .\label{Tt}%
\end{eqnarray}
The first (second) amplitude is related to the longitudinal (transverse) part
of the axial current, respectively. In the chiral limit $w_{L}$ is fixed by
the axial anomaly and the non-renormalization theorem predicts
\cite{Vainshtein:2002nv}
\begin{equation}
w_{L}^{\left(  3,8\right)  }\left(  q^{2}\right)  =2/q^{2}.\label{wL}%
\end{equation}
In the hard limit $q^{2}\gg\Lambda_{\mathrm{QCD}}^{2}$ the perturbative QCD
predicts the asymptotic form $\left(  a=3,8,0\right)  $
\begin{equation}
w_{T}^{\left(  a\right)  }\left(  q^{2}\right)  =\frac{1}{2}w_{L}^{\left(
a\right)  }\left(  q^{2}\right)  =1/q^{2}\label{Was}%
\end{equation}
and (\ref{MVlimit}) transforms into%
\begin{equation}
\mathcal{A}_{\mu_{1}\mu_{2}\mu_{3}\gamma\delta}f^{\gamma\delta}=\frac{8}%
{q_{3}^{2}{\hat{q}}^{2}}\epsilon_{\mu_{1}\mu_{2}\delta\rho}{\hat{q}}^{\delta
}\!\!\left\{  2\,q_{3}^{\rho}q_{3}^{\sigma}\tilde{f}_{\sigma\mu_{3}}+\!\left(
-q_{3}^{2}\tilde{f}_{\mu_{3}}^{\rho}\!+\!q_{3\mu_{3}}q_{3}^{\sigma}\tilde
{f}_{\sigma}^{\rho}\!-\!q_{3}^{\rho}q_{3}^{\sigma}\tilde{f}_{\sigma\mu_{3}%
}\right)  \right\}  +....\label{MVope}%
\end{equation}
It was demonstrated in \cite{Dorokhov:2005pg,Dorokhov:2005hw} that the results
(\ref{Tt}-\ref{Was}) are satisfied within the non-perturbative N$\chi$QM approach.

In the next sections we discuss the basic elements of the N$\chi$QM. Then
within this model we calculate the leading pion-pole contribution to $a_{\mu}%
$, demonstrate how the Melnikov-Vainshtein asymptotics is realized, and
finally discuss on the role of the triangle functions $w_{L\left(  T\right)
}$ for $a_{\mu}$.

\section{Non-local chiral quark model}

To study the LbL contribution to $a_{\mu}$ one can use the framework of the
effective field model of QCD. In the low-momenta domain the effects of the
non-perturbative structure of the QCD vacuum become dominant. Since the
invention of the QCD sum rule method based on the use of the standard OPE, it
is common to parameterize the non-perturbative properties of the QCD vacuum in
terms of infinite towers of the vacuum expectation values of the quark and gluon
operators. From this point of view the nonlocal properties of the QCD vacuum
result from the partial resummation of the infinite series of power
corrections, related to vacuum averages of the quark and gluon operators of growing
dimension, and may be parametrized in terms of the nonlocal vacuum condensates
\cite{Mikhailov:1988nz,Mikhailov:1991pt,Dorokhov:1997iv}. This construction
leads effectively to nonlocal modifications of the propagators and effective
vertices of the quark and gluon fields at small momenta.

An adequate model embodied in this general picture is the instanton model of
the QCD vacuum \cite{Callan:1977gz,Shuryak:1981ff,Diakonov:1985eg} which
describes non-perturbative nonlocal interactions in terms of the effective
action (for a review see \cite{Schafer:1996wv}). Spontaneous breaking of the
chiral symmetry and the dynamical generation of a momentum-dependent quark
mass are naturally explained within the instanton model. The non-singlet and
singlet vector and axial-vector current-current correlators and the vector Adler function
have been calculated in \cite{Dorokhov:2003kf,Dorokhov:2004ze} in the
framework of the effective chiral model with instanton-like nonlocal
quark-quark interaction \cite{Anikin:2000rq,Dorokhov:2003kf}. It was shown, in
particular, that the Weinberg sum rules which are sensitive to the low- as well to
the high-momentum dynamics are satisfied \cite{Broniowski:1999dm,Dorokhov:2003kf}.
In the same model the pion transition form factor normalized by the axial
anomaly has been considered in \cite{Dorokhov:2002iu} for arbitrary photon virtualities.

We start with the nonlocal chirally invariant action\footnote{In the present
work we do not consider extensions of the model to include explicit vector
mesons and the effects of the strange quark mass. Such extensions of the nonlocal model were
considered in \cite{Plant:1997jr,Dorokhov:2004ze} and will be used in
\textit{complete} calculations elsewhere.} which describes the interaction of soft
quark fields \cite{Dorokhov:2000gu},
\begin{eqnarray}
&&  S=\int d^{4}x\ \overline{q}_{I}(x)\left[  i\gamma^{\mu}D_{\mu}%
-m_{f}\right]  q_{I}(x)+\label{Lint}\\
&&  +\frac{1}{2}G_{P}\int d^{4}X\int\prod_{n=1}^{4}d^{4}x_{n}f(x_{n})\left[
\overline{Q}(X-x_{1},X)\right.  \cdot\nonumber\\
&&  \cdot\left.  \Gamma_{P}Q(X,X+x_{3})\overline{Q}(X-x_{2},X)\Gamma
_{P}Q(X,X+x_{4})\right]  ,\nonumber
\end{eqnarray}
where $m_{f}$ is the current quark mass, $D_{\mu}=\partial_{\mu}-iV_{\mu
}\left(  x\right)  -i\gamma_{5}A_{\mu}\left(  x\right)  $ is the covariant
derivative, and the matrix product $\Gamma_{P}\otimes\Gamma_{P}=\left(
1\otimes1+i\gamma_{5}\tau^{a}\otimes i\gamma_{5}\tau^{a}\right)  $ provides
the spin-flavor structure of the interaction. In Eq.~(\ref{Lint})
$\overline{q}_{I}=(\overline{u},\overline{d})$ denotes the flavor doublet
field of dynamically generated quarks, $G_{P}$ is the four-quark coupling
constant, and $\tau^{a}$ are the Pauli isospin matrices. The separable
nonlocal kernel of the interaction described in terms of the form factors
$f(x)$ is motivated by the instanton model of the QCD vacuum, where the
function $f(x)$ may be evaluated. In order to make the nonlocal action
gauge-invariant with respect to the external vector and axial gauge fields
$V_{\mu}^{a}(x)$ and $A_{\mu}^{a}(x)$, we employ in (\ref{Lint}) the
delocalized quark field, $Q(x)$, defined with the help of the Schwinger gauge
phase factors,
\begin{eqnarray}
&&  Q(x,y)=P\exp\left\{  i\int_{x}^{y}dz_{\mu}\left[  V_{\mu}^{a}(z)+\gamma
_{5}A_{\mu}^{a}(z)\right]  T^{a}\right\}  q_{I}(y),\nonumber\\
&&  \overline{Q}(x,y)=Q^{\dagger}(x,y)\gamma^{0}.\label{Qxy}%
\end{eqnarray}
Here $P$ stands for the operator ordering along the integration path, with $y$
denoting the position of the quark and $x$ being an arbitrary reference
point.

The dressed quark propagator, $S(p)$, is found to be\footnote{From here on the
Euclidean metric for the momenta is used.}
\begin{equation}
S^{-1}(p)=i\widehat{p}-M(p^{2}),\label{QuarkProp}%
\end{equation}
with the momentum-dependent quark mass defined as the solution of the gap
equation
\begin{equation}
M(p^{2})=m_{f}+4G_{P}N_{f}N_{c}f^{2}(p^{2})\int\frac{d^{4}k}{\left(
2\pi\right)  ^{4}}f^{2}(k^{2})\frac{M(k^{2})}{k^{2}+M^{2}(k^{2})}.\label{SDEq}%
\end{equation}
The formal solution is expressed as \cite{Bowler:1994ir}
\begin{equation}
M(p^{2})=m_{f}+(M_{q}-m_{f})f^{2}(p^{2}),
\end{equation}
with the constant $M_{q}\equiv M(0)$ determined dynamically from Eq.~(\ref{SDEq}).
The momentum-dependent function $f(p)$ is the normalized four-dimensional
Fourier transform of $f(x)$ given in the coordinate representation. Within the
instanton model $f$ is related to the zero mode solution for the massless
fermion in the field of an instanton. In \cite{Dorokhov:2005pg} it was shown
that in the momentum space the nonlocal function $f\left(  p\right)  $ for
large momenta must decrease faster than any inverse power of $p^{2}$,
\textit{e.g.}, like an exponential, and that this choice corresponds to
calculations in the axial gauge for the quark effective field. In order to
take these effects into account and to make numerics simpler we use for the
nonlocal function the Gaussian form,
\begin{equation}
f(p)=\exp\left(  -p^{2}/\Lambda^{2}\right)  ,\label{MassDyna}%
\end{equation}
where the parameter $\Lambda$ characterizes the nonlocality size of the gluon
vacuum fluctuations and is proportional to the inverse average size of the
instanton in the QCD vacuum.

An important property of the dynamical mass (\ref{SDEq}) is that at low
virtualities its value is close to the constituent mass, while at large
virtualities it goes to the current mass value. This property is crucial in
obtaining the correct, consistent with the OPE, QCD behavior of different
correlators at large momentum transfers
\cite{Dorokhov:2003kf,Dorokhov:2004ze,Dorokhov:2005pg}. The nonlocal chiral
quark model can be viewed as an approximation to the large-$N_{c}$ QCD, where
the only (effective) interaction terms, retained after integrating out the
high-frequency modes of the quark and gluon fields down to the nonlocality
scale $\Lambda$ where the spontaneous chiral symmetry breaking occurs, are
those which can be cast in the form of the four-fermion operators
(\ref{Lint}). The parameters of the model are then the nonlocality scale
$\Lambda$, the four-fermion coupling constant $G_{P}$, and the current quark
masses $m_{f}$.

The quark-antiquark scattering matrix in the pseudoscalar channel is found
from the Bethe-Salpeter equation as
\begin{equation}
\widehat{T}_{P}(q^{2})=\frac{G_{P}}{1-G_{P}J_{PP}(q^{2})}, \label{ScattMatr}%
\end{equation}
with the polarization operator in the pseudoscalar channel equal to
\begin{equation}
J_{PP}(q^{2})=\int\frac{d^{4}k}{\left(  2\pi\right)  ^{4}}f^{2}\left(
k\right)  f^{2}\left(  k+q\right)  Tr\left[  S(k)\gamma_{5}S\left(
k+q\right)  \gamma_{5}\right]  . \label{J}%
\end{equation}
The position of the pion state is determined as the pole of the scattering
matrix
\begin{equation}
\left.  \det(1-G_{P}J_{PP}(q^{2}))\right\vert _{q^{2}=-m_{\pi}^{2}}=0.
\label{PoleEq}%
\end{equation}
The quark-pion vertex identified with the residue of the scattering matrix is
$\left(  k^{\prime}=k+q\right)  $
\begin{equation}
\Gamma_{\pi}^{a}\left(  k,k^{\prime}\right)  =g_{\pi qq}i\gamma_{5}\tau
^{a}f(k)f(k^{\prime}) \label{PiVertex}%
\end{equation}
with the quark-pion coupling equal to
\begin{equation}
g_{\pi q}^{-2}=-\left.  \frac{dJ_{PP}\left(  q^{2}\right)  }{dq^{2}%
}\right\vert _{q^{2}=-m_{\pi}^{2}}, \label{gM}%
\end{equation}
where $m_{\mathrm{\pi}}$ is the physical mass of the pion. In the chiral limit
the quark-pion coupling, $g_{\pi q}$, and the pion decay constant, $f_{\pi}$,
are connected by the Goldberger-Treiman relation, $g_{\pi}=M_{q}/f_{\pi},$
which is verified to be valid in N$\chi$QM, as requested by the chiral symmetry.

\section{Conserved vector and axial-vector currents}

\label{IV}

The vector vertex following from the model (\ref{Lint}) is
\begin{equation}
\Gamma_{\mu}(k,k^{\prime})=\,\hat{Q}\left[  \gamma_{\mu}+(k+k^{\prime})_{\mu
}M^{(1)}(k,k^{\prime})\right]  , \label{GV}%
\end{equation}
where $\hat{Q}$ is the diagonal matrix of the quark electric charges,
$M^{(1)}(k,k^{\prime})$ is the finite-difference derivative of the dynamical
quark mass, $k$ $(k^{\prime})$ is the incoming (outgoing) momentum of the
quark, and $q=k^{\prime}-k$ is the momentum corresponding to the current. The
finite-difference derivative of an arbitrary function $F$ is defined as
\begin{equation}
F^{(1)}(k,k^{\prime})=\frac{F(k^{\prime})-F(k)}{k^{\prime2}-k^{2}}.
\label{FDD}%
\end{equation}
The full axial vertex corresponding to the conserved axial-vector current is
obtained after the resummation of quark-loop chain that results in the form of
the term proportional to the pion propagator \cite{Anikin:2000rq}
\begin{eqnarray}
&\Gamma_{\mu}^{a,5}(k,k^{\prime})  &  =\left[  \gamma_{\mu}\gamma_{5}%
+\Delta\Gamma_{\mu}^{5}(k,k^{\prime})\right]  \lambda_{a},\nonumber\\
&\Delta\Gamma_{\mu}^{5}(k,k^{\prime})  &  =2\gamma_{5}\frac{q_{\mu}}{q^{2}%
}f(k)f(k^{\prime})\left[  J_{AP}\left(  0\right)  -\frac{m_{f}G_{P}%
J_{P}\left(  q^{2}\right)  }{1-G_{P}J_{PP}\left(  q^{2}\right)  }\right]
+\gamma_{5}(k+k^{\prime})_{\mu}J_{AP}\left(  0\right)  \frac{\left(
f(k^{\prime})-f\left(  k\right)  \right)  ^{2}}{k^{\prime2}-k^{2}},
\label{GAtot}%
\end{eqnarray}
where we have introduced the notation
\begin{equation}
J_{P}(q^{2})=\int\frac{d^{4}k}{\left(  2\pi\right)  ^{4}}f\left(  k\right)
f\left(  k+q\right)  Tr\left[  S(k)\gamma_{5}S\left(  k+q\right)  \gamma
_{5}\right]  , \label{JP}%
\end{equation}%
\begin{equation}
J_{AP}(q^{2})=4N_{c}N_{f}\int\frac{d^{4}l}{\left(  2\pi\right)  ^{4}}%
\frac{M\left(  l\right)  }{D\left(  l\right)  }\sqrt{M\left(  l+q\right)
M\left(  l\right)  }. \label{JAP}%
\end{equation}
In the chiral limit of $m_{f}=0$ one gets instead%
\begin{equation}
\Delta\Gamma_{\mu}^{5}(k,k^{\prime})=\gamma_{5}\frac{q_{\mu}}{q^{2}}\left[
M(k^{\prime})+M\left(  k\right)  \right]  +\gamma_{5}\left[  \frac{q_{\mu}%
}{q^{2}}-\frac{(k+k^{\prime})_{\mu}}{k^{\prime2}-k^{2}}\right]  M_{q}\left(
f(k^{\prime})-f\left(  k\right)  \right)  ^{2}. \label{dGAchiral}%
\end{equation}
The axial-vector vertex (\ref{GAtot}) has a pole at
\begin{equation}
q^{2}=-m_{\pi}^{2}=m_{f}\left\langle \overline{q}q\right\rangle /f_{\pi}^{2}%
\end{equation}
where the Goldberger-Treiman relation and the definition of the quark
condensate has been used. The pole is related to the denominator
$1-G_{P}J_{PP}\left(  q^{2}\right)  $ in Eq. (\ref{GAtot}), while $q^{2}$ in
denominator is compensated by zero from the square bracket in the limit
$q^{2}\rightarrow0.$ This compensation follows from expansion of $J(q^{2})$
functions near zero momentum
\begin{eqnarray}
&J_{PP}(q^{2})  &  =G_{P}^{-1}-m_{f}\left\langle \overline{q}q\right\rangle
M_{q}^{-2}-q^{2}g_{\pi q}^{-2}+O\left(  q^{4}\right)  ,\qquad\label{LowQ}\\
&J_{AP}(q^{2}=0)  &  =M_{q},\qquad J_{P}(q^{2}=0)=\left\langle \overline
{q}q\right\rangle M_{q}^{-1}.\nonumber
\end{eqnarray}
It follows from this expansion that near pole position, $q^{2}=-m_{\pi}^{2}$,
one has for the factor in (\ref{PionPole})
\begin{equation}
\frac{G_{P}\ }{g_{\pi q}^{2}\left(  1-G_{P}J_{pp}\left(  q_{2}^{2}\right)
\right)  }=\frac{1}{q^{2}+m_{\pi}^{2}}+.... \label{Pole}%
\end{equation}
However, at large $q^{2}$ this factor goes to the constant $G_{P}g_{\pi
q}^{-2}$ instead of decay like $q^{-2}$ \ as it would be predicted by the simple
pole approximation. In the chiral limit $m_{f}=0$ the second structure in
square brackets in Eq. (\ref{GAtot}) disappears and the pole moves to zero. It
should be stressed that the form of the interaction vertices (\ref{GV}%
,\ref{GAtot}) is consistent with the chiral Ward-Takahashi identities.

Within N$\chi$QM the full singlet axial-vector vertex including local and
nonlocal pieces is given by \cite{Dorokhov:2003kf}
\begin{eqnarray}
&\Gamma_{\mu}^{0,5}(k,k^{\prime}) &  =\gamma_{\mu}\gamma_{5}+\Delta\Gamma_{\mu
}^{0,5}(k,k^{\prime}),\label{G50}\\
&\Delta\Gamma_{\mu}^{0,5}(k,k^{\prime}) &  =\gamma_{5}\left[  (k+k^{\prime
})_{\mu}M_{q}\frac{\left(  f\left(  k^{\prime}\right)  -f\left(  k\right)
\right)  ^{2}}{k^{\prime2}-k^{2}}+\frac{q_{\mu}}{q^{2}}2M_{q}f\left(
k^{\prime}\right)  f\left(  k\right)  \frac{G_{P}^{0}}{G_{P}}\frac
{1-G_{P}J_{PP}(q^{2})}{1-G_{P}^{0}J_{PP}(q^{2})}\right]  ,\nonumber
\end{eqnarray}
where $G_{P}^{0}$ is the four-quark coupling for the singlet channel. As it
follows from expansion (\ref{LowQ}) the singlet current (\ref{G50}) does not
contain massless pole due to presence of the $U_{A}\left(  1\right)  $
anomaly. Instead, the singlet current develops a pole at the $\eta^{\prime}-$
meson mass
\begin{equation}
1-G_{P}^{0}J_{PP}(q^{2}=-m_{\eta^{\prime}}^{2})=0,\label{Eta1}%
\end{equation}
thus solving the $U_{A}(1)$ problem.

The parameters of the model are fixed in a way typical for effective
low-energy quark models. One usually fits the pion decay constant, $f_{\pi}$,
and the pion mass (\ref{PoleEq}) to their experimental values. In the chiral
limit the decay constant reduces to $f_{0,\pi}=86$~\textrm{MeV}
\cite{Gasser:1983yg}. In N$\chi$QM $f_{0,\pi}$ is expressed as
\cite{Diakonov:1985eg}
\begin{equation}
f_{0,\pi}^{2}=\frac{N_{c}}{4\pi^{2}}\int\limits_{0}^{\infty}dk^{2}\ k^{2}%
\frac{M^{2}(k^{2})-k^{2}M(k^{2})M^{\prime}(k^{2})+k^{4}M^{\prime}(k^{2})^{2}%
}{\left(  k^{2}+M^{2}(k^{2})\right)  ^{2}}, \label{Fpi2_M}%
\end{equation}
where the primes mean the derivatives with respect to $k^{2}$, \emph{i.e.}
$M^{\prime}(k^{2})=dM(k^{2})/dk^{2}$, \textit{etc}. The described fitting
procedure introduces two relations among the three model parameters. We use
the following values of the model parameters fixed in
Ref.~\cite{Dorokhov:2004ze}:
\begin{equation}
M_{q}=0.24~\mathrm{GeV,}\qquad\Lambda_{P}=1.11~\mathrm{GeV,\quad}%
m_{f}=8~\mathrm{MeV}. \label{param1}%
\end{equation}
To test the sensitivity of the results on the choice of parameters we also use
the set \cite{Dorokhov:2002iu}
\begin{equation}
M_{q}=0.35~\mathrm{GeV,}\qquad\Lambda_{P}=1.2~\mathrm{GeV,\quad}%
m_{f}=12~\mathrm{MeV}. \label{param2}%
\end{equation}
Both sets reproduce the low-energy observables with an acceptable accuracy of the order of 10\%.
The finite current quark mass in (\ref{param1}) and (\ref{param2}) is fixed by the physical value of
pion mass. It also increases the pion decay constant to its experimental value,
$f_{\pi}=92$~\textrm{MeV}, within a 2 MeV accuracy.

\begin{figure}[th]
\includegraphics[width=15cm]{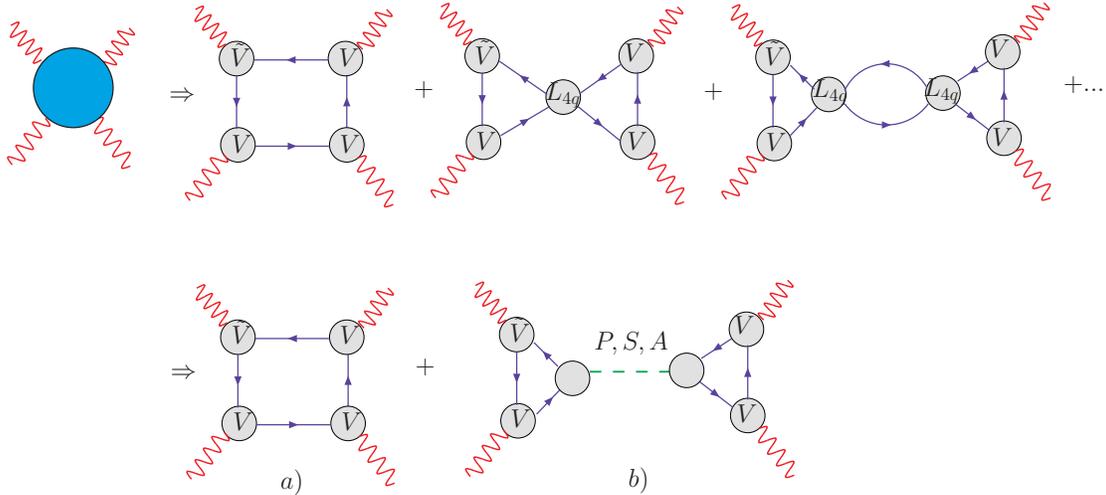}\caption{(Color online) The diagrammatic
representation of the LbL scattering amplitude within the effective four-quark
model in the leading $1/N_{c}$ approximation. It consists of the box diagram
plus the iteration of the four-quark interaction term via the quark loop. The
iterative terms sum up into a two-point meson correlator in the given channel
(pseudoscalar, scalar, or axial-vector). All quark lines and the vertices are
dressed, as calculated within N$\chi$QM. There are also a number of contact
terms inherent to nonlocal models, not shown in the figure. }%
\label{4by4}%
\end{figure}

The triangle $PVV$ amplitude corresponding to the process $\gamma^{\ast}%
\gamma^{\ast}\rightarrow\pi_{0}^{\ast}$
\begin{equation}
A\left(  \gamma^{\ast}\left(  q_{1},\epsilon_{1}\right)  \gamma^{\ast}\left(
q_{2},\epsilon_{2}\right)  \rightarrow\pi^{0}\left(  p\right)  \right)
=-ie^{2}\varepsilon_{\mu\nu\rho\sigma}\epsilon_{1}^{\mu}\epsilon_{2}^{\nu
}q_{1}^{\rho}q_{2}^{\sigma}\mathcal{F}_{\pi_{0}^{\ast}\gamma^{\ast}%
\gamma^{\ast}}(q_{3}^{2};q_{1}^{2},q_{2}^{2}),\label{AF}%
\end{equation}
has been constructed and discussed in \cite{Dorokhov:2002iu}\footnote{In
\cite{Dorokhov:2002iu} the amplitude $\mathcal{F}_{\pi_{0}^{\ast}\gamma
\ast\gamma^{\ast}}$ is denoted as $M_{\pi_{0}}$.}. In (\ref{AF}) $\epsilon
_{i}^{\mu}$ and $q_{i}$ $(i=1,2)$ are the photon polarization vectors and
momenta, while $q_{3}=q_{1}+q_{2}$ is the pion momentum. In N$\chi$QM one
finds the contribution of the triangle diagram to the invariant amplitude as
\begin{eqnarray}
A\left(  \gamma_{1}^{\ast}\gamma_{2}^{\ast}\rightarrow\pi_{0}^{\ast}\right)
&&  =-ie^{2}\frac{N_{c}}{f_{\pi}}Tr\left[  \widehat{Q}^{2}\tau^{3}\right]
\epsilon_{1}^{\mu}\epsilon_{2}^{\nu}\int\frac{d^{4}k}{(2\pi)^{4}}%
Tr[\Gamma_{\pi}^{a}\left(  k_{+},k_{-}\right)  S(k_{-})\Gamma_{\mu}%
(k_{-},k_{3})S\left(  k_{3}\right)  \Gamma_{\nu}(k_{3},k_{+})S(k_{+}%
)]\label{MPiGG}\\
&&  +\left(  q_{1}\leftrightarrow q_{2};\epsilon_{1}\leftrightarrow\epsilon
_{2}\right)  .\nonumber
\end{eqnarray}
where $q=q_{1}-q_{2},$ $k_{\pm}=k\pm q_{3}/2,~k_{3}=k-q/2$. In the chiral
limit $\left(  q_{3}^{2}=m_{\pi}^{2}=0\right)  $ with both photons real
$\left(  q_{i}^{2}=0\right)  $ one finds the normalization by the axial anomaly
(within the nonlocal models it was proven in \cite{Plant:1997jr})%
\begin{equation}
\mathcal{F}_{\pi_{0}^{\ast}\gamma^{\ast}\gamma^{\ast}}\left(  0;0,0\right)
=\frac{N_{c}}{12\pi^{2}f_{\pi}}.\label{ChAn}%
\end{equation}
The functions defined in (\ref{ScattMatr}), (\ref{J}), (\ref{AF}),
(\ref{MPiGG}) are used in the evaluation of (\ref{PionPole}).
Within the effective model the light-by-light amlitude is given by diagrams of Fig. \ref{4by4}.

\section{Numerical results for the pion pole contribution to LbL}

\begin{figure}[th]
\includegraphics[width=7.5cm]{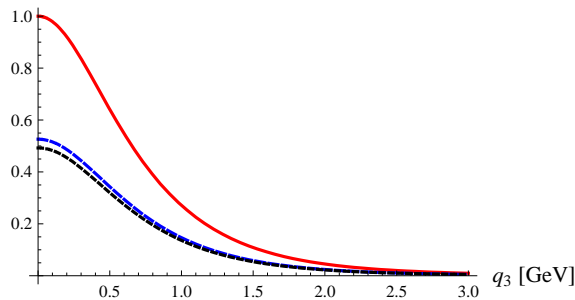}\caption{(Color online) The dependence
of the pion transition form factor $\mathcal{F}_{\pi_{0}^{\ast}\gamma^{\ast
}\gamma^{\ast}}(q_{3}^{2};q_{1}^{2},q_{2}^{2})/\mathcal{F}_{\pi_{0}^{\ast
}\gamma^{\ast}\gamma^{\ast}}(0;0,0)$ on the off-shellness of the pion,
$|q_{3}|$, for $q_{1}^{2}=q_{2}^{2}=0$ (top curve), $q_{1}^{2}%
=(700~\mathrm{MeV})^{2}$, $q_{2}^{2}=0$ (middle curve), and for $q_{1}%
^{2}=q_{2}^{2}=(700~\mathrm{MeV})^{2}/2$ (bottom curve). We note a strong
dependence on $|q_{3}|$ (all momenta are Euclidean). }%
\label{fig:ffc}%
\end{figure}

We first investigate the role of the off-shellness in the pion transition form
factor $\mathcal{F}_{\pi_{0}^{\ast}\gamma^{\ast}\gamma^{\ast}}(q_{3}^{2}%
;q_{1}^{2},q_{2}^{2})$, which is the {important} element considered in this work. In
Fig.~\ref{fig:ffc} we show how $\mathcal{F}_{\pi_{0}^{\ast}\gamma^{\ast}%
\gamma^{\ast}}(q_{3}^{2};q_{1}^{2},q_{2}^{2})/\mathcal{F}_{\pi_{0}^{\ast
}\gamma^{\ast}\gamma^{\ast}}(0;0,0)$ depends on $q_{3}$ for fixed values of
$q_{1}$ and $q_{2}$. For all cases of $q_{1}$ and $q_{2}$ we note a strong
dependence on $q_{3}$, which for large values of this momentum becomes
exponential, $\sim\exp(-\mathrm{const}\cdot q_{3}^{2})$. This strong
dependence on the off-shellness, related to the Gaussian  form of the nonlocal
quark-pion vertex (\ref{PiVertex}), could in principle lead to a significant
reduction of the pion pole contribution to LbL. As we shall see below, this is
not the case, since the effective support for the integrand of $a_{\mu
}^{\mathrm{LbL,\pi}_{0}}$ is localized at moderate values of $q_{3}^{2}$ where
the form factor suppression is not so strong. Furthermore, the factor $\left(
1-G_{P}J_{pp}\left(  q_{3}^{2}\right)  \right)  ^{-1}$ in Eq. (\ref{PionPole})
at large $q_{3}^{2}$ goes to one, but not to $q_{3}^{-2}$ as in the case of
the single-pole approximation $\left(  q_{3}^{2}-m_{\pi}^{2}\right)  ^{-1}$.

To obtain $a_{\mu}^{\mathrm{LbL,\pi}_{0}}$, one has to convolute the
three-point vertex functions discussed above in the two-loop integral of
Eq.~(\ref{PionPole}). Since the multidimensional integrand is a regular
function, this task can be accomplished straightforwardly with the Monte Carlo
integration technique. We use the routine \texttt{VEGAS} \cite{vegas}, which
implements the adaptive Monte Carlo algorithm. The
eight-dimensional\footnote{Through the use of the rotational symmetry one may
reduce the integral down to five dimensions, but from the view point of the
Monte Carlo integration this is irrelevant for the performance.} integral in
(\ref{PionPole}), carried over the the two muon loop momenta, has been done
numerically in the Euclidean space.

A convenient and important trick, allowing to avoid the numerical cut-offs in
the momentum integration, is to map the momentum variables via a conformal
transformation into the range $[0,1)$. Explicitly, we use
\begin{equation}
\xi_{i}=\frac{p_{i}^2}{p_{i}^2+a},
\end{equation}
with $p_{i}$ denoting the Euclidean momentum and $a=1~{\rm GeV}^2$. We have tested the accuracy of
our code by reproducing very accurately various model results listed in
Ref.~\cite{Knecht:2001qf}.

The result of the calculation in N$\chi$QM with parameters (\ref{param1}) is
\begin{equation}
a_{\mu}^{\mathrm{LbL,\pi}_{0}}=6.27\cdot10^{-10}. \label{resu1}%
\end{equation}
while with parameters (\ref{param2}) we get
\begin{equation}
a_{\mu}^{\mathrm{LbL,\pi}_{0}}=6.68\cdot10^{-10}. \label{resu2}%
\end{equation}
Similarly to other models, our result is dominated by the term proportional to
the $T_{1}$ structure in Eq.~(\ref{PionPole}), which yields 97\% of the total
of (\ref{resu1},\ref{resu2}).

\begin{figure}[th]
\includegraphics[width=7.5cm]{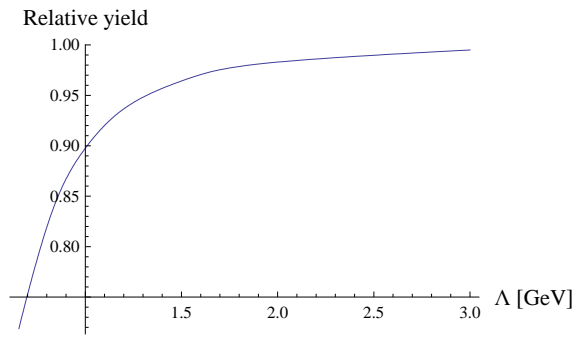}\caption{(Color online) The relative yield of the full result
for $a_{\mu}^{\mathrm{LbL,\pi}_{0}}$ for model parameters  (\ref{param2}),
plotted as a function of the Euclidean momentum cut-off, $\Lambda$.
\label{fig:ry}}%
\end{figure}

In order to illustrate the convergence of the result with the increasing upper limit of
the momentum integration, here denoted by $\Lambda$, we show in Fig. \ref{fig:ry} the
relative yield of the full result for $a_{\mu}^{\mathrm{LbL,\pi}_{0}}$ with model
parameters  (\ref{param2}), plotted as a function of the Euclidean momentum cut-off,
$\Lambda$ ($|p_i| \le \Lambda$). For $\Lambda \to \infty$ we obtain, by definition, the
full result. For $\Lambda=1$ GeV about 90\% of the full result is obtained. Thus the
result is dominated by the soft physics, as requested of the effective low-energy model.
These findings are in agreement with that obtained in
\cite{Knecht:2001qf,Bijnens:2007pz}.

Finally, we note that in the extended version of the considered model  the vector and axial-vector
couplings appear. However, as it was discussed in \cite{Dorokhov:2002iu} within the
nonlocal model, the inclusion of these couplings leads to very minor changes of the pion
transition form factor. In particular, it does not change the normalization of the form
factor at zero momentum and contributes small corrections to the leading-order
asymptotics. The latter is due to the relatively large mass of the rho meson.

\section{Comparison to other approaches}

In this section we compare our calculation of the pion-pole light-by-light
contribution to $a_{\mu}$ with the results obtained in earlier calculations
\cite{Bijnens:2001cq,Hayakawa:1997rq,Knecht:2001qf}
\begin{eqnarray}
&a_{\mu}^{\pi^{0}\mathrm{{,LbL}}}  &  =5.6\cdot10^{-10}\ \ \ \ \ \ \ \ \ \ \ \   [11]
,\label{lpiB}\\
&a_{\mu}^{\pi^{0}\mathrm{{,LbL}}}  &  =5.6\cdot10^{-10}\ \ \ \ \ \ \ \ \ \ \ \   [13]
,\label{lpiH}\\
&a_{\mu}^{\pi^{0}\mathrm{{,LbL}}}  &  =5.8(1.0)\cdot10^{-10}\ \ \ \ \ \ [15].
\label{lpiM}\end{eqnarray}
These calculations are based on the usage of different parameterizations of
the $\pi\gamma^{\ast}\gamma^{\ast}$ vertex satisfying the CLEO data on the
pion transition form factor $\pi\to\gamma\gamma^{\ast}$ and OPE constraints.
Thus, in \cite{Knecht:2001qf}
the phenomenological form factor from the generalized vector-meson dominance (VMD)
was used (the Knecht-Nyffeler (KN) model),
\begin{equation}
F_{\pi\gamma^{\ast}\gamma^{\ast}}^{\mathrm{gVMD}}\left(  s,t\right)
=\frac{f_{\pi}}{3}\frac{\left(  s+t\right)  st-h_{2}st+h_{5}\left(
s+t\right)  +M_{V}^{4}M_{V_{1}}^{4}h_{7}}{\left(  M_{V}^{2}+s\right)  \left(
M_{V}^{2}+t\right)  \left(  M_{V_{1}}^{2}+s\right)  \left(  M_{V_{1}}%
^{2}+t\right)  }, \label{gVMD}%
\end{equation}
with the parameters $M_{V}=769$ MeV, $M_{V_{1}}=1465$ MeV, $h_{5}=6.93$
GeV$^{4}$, $h_{7}=N_{c}/\left(  4\pi^{2}f_{\pi}^{2}\right)  $. The error in
(\ref{lpiM}) is due to the uncertainty of the parameter $h_{2}$ taken from the
interval $h_{2}\in\lbrack10,-10]$ GeV$^{2}$. However, the function
(\ref{gVMD}) depends on two kinematic variables instead of three required by
(\ref{PionPole}). We test the effect of this approximation. Namely, in
(\ref{MPiGG}) we have removed the dependence on the pion virtuality and have
taken the meson propagator in (\ref{PionPole}) in the single-pole
approximation $1/\left(  p^{2}-m_{\pi}^{2}\right)  $.
In the N$\chi$QM we obtain numerically for the parameter set
(\ref{param1})
\begin{equation}
a_{\mu}^{\mathrm{LbL,\pi}_{0}}=7.33\cdot10^{-10}%
\;\;\;\;(\mathrm{{no~off-shellness~effects}).}%
\end{equation}
This result should be compared to the full calculation yielding (\ref{resu1}).
Thus, as already stated, the inclusion of the dependence on the pion
virtuality reduces the result by about 15\%.
{A similar statement that the influence of the pion off-shellness has only a
small effect was presented in \cite{Knecht:2001qf}. In that work different
form factors, including the point-like pion coupling and the form factors used in the
extended Nambu--Jona-Lasinio (ENJL) model of
\cite{Bijnens:2001cq,Bijnens:1995xf}, were tested. The stability of the results
was demonstrated.}

Recently, there was an attempt by Jegerlehner and Nyffeler \cite{JN08} to
improve the KN model by taking into account the full kinematic dependence of the
pion-photon vertex through the use of the parametrization for $F_{\pi^{\ast}\gamma
^{\ast}\gamma^{\ast}}^{\mathrm{gVMD}}\left(  s,t,u\right)  $ suggested in
\cite{Knecht:2001xc}. This yields for the sum of the $\pi,\eta,\eta^{\prime}$
contributions $a_{\mu}^{\mathrm{LbL,\pi}_{0},\eta,\eta^{\prime}}%
=9.66\pm4.5\cdot10^{-10}$. As a result an enhancement of the contribution in
comparison with KN result $a_{\mu}^{\mathrm{LbL,\pi}_{0},\eta,\eta^{\prime}%
}=8.3\pm1.2\cdot10^{-10}$ was obtained.
{We believe that this is a result of neglecting the dressing effects in the
parametrization constructed in \cite{Knecht:2001xc}. In particular, this
parametrization reproduces the short-distance behavior of the three-point correlator of
the pseudoscalar and vector currents \cite{Knecht:2001xc}. For these currents the local
$\gamma_5$ and $\gamma_\mu$ couplings were used, correspondingly. Because of the local
character of these vertices the correlator has a slow power-like behaviour at large
momenta. However, this correlator does not correspond to the triangle quark diagram
connecting the virtual pion with the photons. The pion has its own hadronic form factor, which
means that the $\gamma_5$ vertex is dressed, as it corresponds to the nonlocal form factor
(\ref{PiVertex}) (see the similar results in the Schwinger-Dyson approach
\cite{Maris:1998hc}). As a result the large-momentum properties of the calculation
change drastically.}

Similar problems are encountered in calculations based on the usage of the constant
constituent quark mass (the local models) \cite{Pivovarov:2001mw,Bartos:2001pg}.
Typically, these models predict larger numbers for the hadronic contribution due
to incorrect form factors, artificially enhanced with respect to the OPE asymptotics.

We have to note that our approach is closest to the formalism of the work
\cite{Bijnens:1995xf} (ENJL model) in the sense of the use of techniques of
the effective models. However, the physical grounds are quite different. The ENJL
model is in fact a generalization of the vector meson dominance model, while
N$\chi$QM is based on the fundamental property of the QCD vacuum, namely
the nonlocality of the vacuum fluctuations of the gluon field . Unfortunately, the region of
applicability of the ENJL model is restricted to momenta much lower than
1~GeV. For example, we are not able to prove the properties of the triangle
diagram (\ref{Tt}-\ref{Was}) by using the $VVA$ functions found in the
ENJL model \cite{Bijnens:1995xf}. The inconsistency of the Adler function
found within ENJL with the OPE was noted earlier in \cite{Dorokhov:2004ze}. The
high-momentum region was then modeled in \cite{Bijnens:1995xf} by different
parameterizations, which may be quite risky. One of us (AD) has already
pointed out in \cite{Dorokhov:2004ze} that a possible reason for the failure
of ENJL at large momenta is due to the fact that this model is based on the
single resonance approximation, while nonlocalities, inherent in N$\chi$QM,
effectively take into account an infinite number of resonances according to
the quark-hadron duality principle. One of the results that is observed in
\cite{Bijnens:1995xf} is that there are large cancellations of different
contributions. This might be a model-dependent statement that has to be
checked in future \textit{complete} calculations within the nonlocal chiral quark
model with instanton-like interactions.

\section{The Melnikov-Vainshtein constraint in N$\chi$QM}

The sole contribution of the pion pole discussed in the previous sections and
also used in \cite{Bijnens:2001cq} and \cite{Knecht:2001qf} does not satisfy
the Melnikov-Vainshtein (MV) asymptotics (\ref{MVope}) discussed in
Sect.~\ref{I}. In the present section we show within the N$\chi$QM model that
it is the quark box diagram of the light-by-light scattering that provides the
correct asymptotics (Fig. \ref{4by4}a). \begin{figure}[th]
\hspace*{0.1cm} \begin{minipage}{6.5cm}
\vspace*{0.5cm} \epsfxsize=6cm \epsfysize=5cm \centerline{\epsfbox
{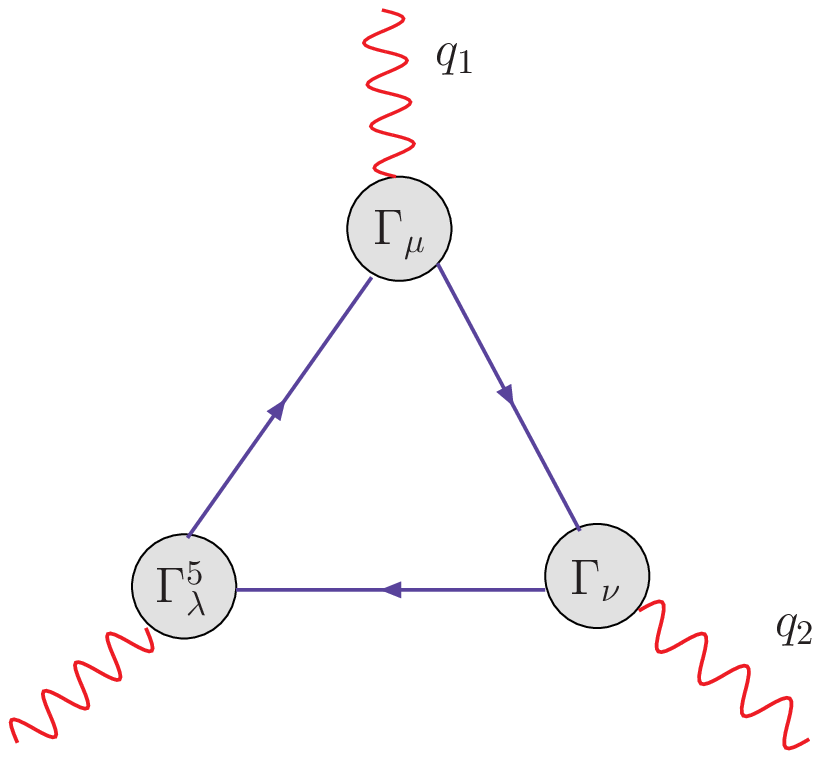}}
\caption[dummy0]{(Color online) Diagrammatic representation of the triangle diagram in the instanton model
with dressed quark lines and full quark-current vertices\label{TriAn}}
\end{minipage}\hspace*{0.5cm} \begin{minipage}{6.5cm}
\vspace*{0.5cm} \epsfxsize=6cm \epsfysize=5cm \centerline{\epsfbox{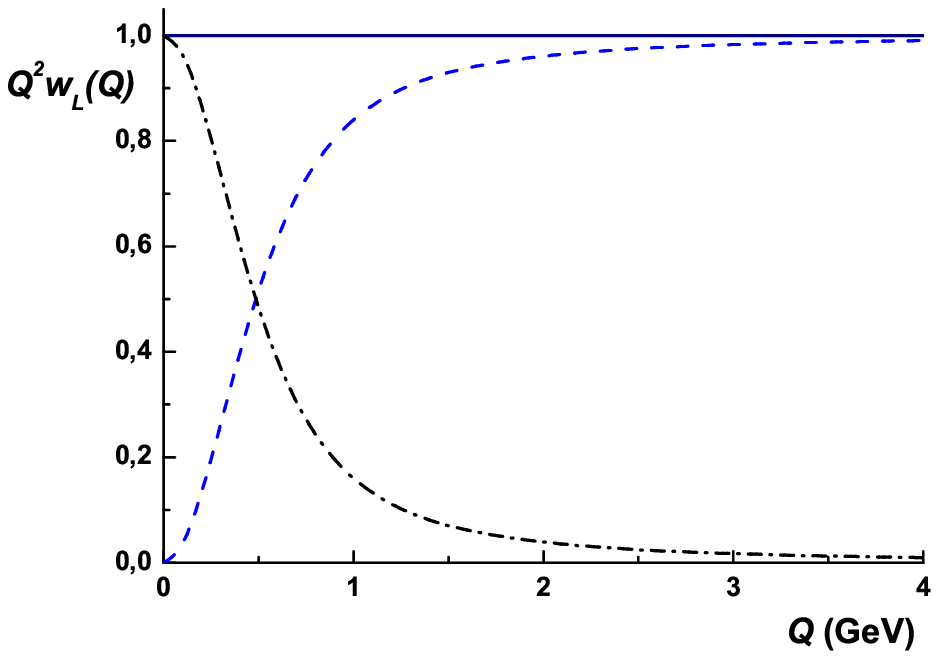}}
\caption[dummy0]{(Color online) The normalized nonsinglet invariant function $w_L$ constrained by the Adler-Bell-Jackiw anomaly following
from the triangle diagram. The dashed line denotes the local part and the dash-dotted line
the nonlocal part.\label{WL}}
\end{minipage}\end{figure}To this end we first recall the properties of the
non-diagonal correlator of the vector current and the axial-vector current in
the external electromagnetic field ($VA\widetilde{V}$) (\ref{Triangle}), as
derived in \cite{Dorokhov:2005pg,Dorokhov:2005hw} within N$\chi$QM. In this
model the $VA\widetilde{V}$ correlator is defined by (Fig. \ref{TriAn})
\begin{eqnarray}
&&  T_{\mu_{3}\rho}^{a}=-2eN_{c}Tr\left[  \widehat{Q}^{2}\lambda^{a}\right]
\epsilon_{4}^{\nu}\int\frac{d^{4}k}{\left(  2\pi\right)  ^{4}}Tr\left[
\Gamma_{\mu_{3}}\left(  k+q_{3},k\right)  S\left(  k+q_{3}\right)  \right.
\cdot\nonumber\\
&&  \left.  \cdot\Gamma_{\rho}^{a,5}\left(  k+q_{3},k-q_{4}\right)  S\left(
k-q_{4}\right)  \widetilde{\Gamma}_{\nu}\left(  k,k-q_{4}\right)  S\left(
k\right)  \right]  ,\label{Tncqm}%
\end{eqnarray}
where for definiteness we consider $q_{4}$ as a soft momentum. The quark
propagator, the vector vertex, and the nonsinglet and singlet axial-vector
vertices are given in (\ref{QuarkProp}), (\ref{GV}), (\ref{GAtot}), and
(\ref{G50}), respectively. The structure of the vector vertices guarantees
that the amplitude is transverse with respect to the vector indices
\begin{equation}
T_{\mu_{3}\rho}^{a}q_{3}^{\mu_{3}}=0,
\end{equation}
while the Lorentz structure of the amplitude is given by (\ref{Tt}). It is
convenient to separate the correlator into three pieces:
\begin{eqnarray}
&T_{\mu_{3}\rho}^{a,\mathrm{Loc}} &  =-2eN_{c}Tr\left[  \widehat{Q}^{2}%
\lambda^{a}\right]  \epsilon_{4}^{\nu}\int\frac{d^{4}k}{\left(  2\pi\right)
^{4}}Tr\left[  \gamma_{\mu_{3}}S\left(  k+q_{3}\right)  \gamma_{\rho}%
\gamma_{5}S\left(  k-q_{4}\right)  \gamma_{\nu}S\left(  k\right)  \right]
,\label{Tloc}\\
&T_{\mu_{3}\rho}^{a,\mathrm{Pole}} &  =-2eN_{c}Tr\left[  \widehat{Q}^{2}%
\lambda^{a}\right]  \epsilon_{4}^{\nu}\int\frac{d^{4}k}{\left(  2\pi\right)
^{4}}Tr\left[  \Gamma_{\mu_{3}}\left(  k+q_{3},k\right)  S\left(
k+q_{3}\right)  \Delta\Gamma_{\rho}^{5}\left(  k+q_{3},k-q_{4}\right)
S\left(  k-q_{4}\right)  \widetilde{\Gamma}_{\nu}\left(  k,k-q_{4}\right)
S\left(  k\right)  \right]  ,\label{Tpole}\\
&T_{\mu_{3}\rho}^{a,\mathrm{\operatorname{Re}st}} &  =T_{\mu_{3}\rho}%
^{a}-T_{\mu_{3}\rho}^{a,\mathrm{Loc}}-T_{\mu_{3}\rho}^{a,\mathrm{Pole}%
}.\label{Trest}%
\end{eqnarray}
In Fig.~\ref{WL} it is shown (for details see \cite{Dorokhov:2005pg}) how the
different terms saturate the anomalous nonsinglet amplitude $w_{L}$
(\ref{wL}). The part $T_{\mu_{3}\rho}^{a,\mathrm{Loc}}$, where all vertices
are local, saturates the anomaly at large momenta, which is just the property
of the quark triangle diagram (dashed line in Fig. \ref{WL})%
\begin{equation}
\left.  T_{\mu_{3}\rho}^{a,\mathrm{Loc}}\right\vert _{\substack{q_{4}%
\mathrm{\ soft}\\q_{3}\gg1\mathrm{\ GeV}}}\rightarrow T_{\mu_{3}\rho}%
^{a}.\label{TriAs}%
\end{equation}
This result (\ref{TriAs}) is independent of the channel and the relations
(\ref{Was}) become true (see \cite{Dorokhov:2005pg,Dorokhov:2005hw} for details).

It is the pole term (\ref{Tpole}) which dominates the low-momentum behavior
of the amplitude sensitive to the flavor structure of the axial
current. In the nonsinglet channel at zero momentum the anomaly is saturated
by the massless pion pole contribution (\ref{Tpole}) (dash-dotted line in Fig.
\ref{WL}). The remaining part $T_{\mu_{3}\rho}^{a,\mathrm{\operatorname{Re}%
st}}$, which vanishes at zero and infinite momenta and is numerically small
everywhere, accomplishes the exact saturation to the correct value at all
momenta. On the other hand, in the singlet case due to different structure of
the singlet current (\ref{G50}) the pole contribution (\ref{Tpole}) is almost
completely compensated and the local part (\ref{Tloc}) dominates the amplitude
at all momenta \cite{Dorokhov:2005hw}.

With this background in mind let us now consider the LbL scattering amplitude
coming from the box diagram with dynamical quarks ($q_{4}$ is a soft momentum)%
\begin{eqnarray}
\mathcal{A}_{\mu_{1}\mu_{2}\mu_{3}\gamma\delta}^{\mathrm{Box}}f^{\gamma\delta}
&&  =\epsilon_{4}^{\mu_{4}}\int\frac{d^{4}k}{\pi^{2}}Tr\left[  S\left(
k\right)  \Gamma_{\mu_{4}}\left(  k,k+q_{4}\right)  S\left(  k+q_{4}\right)
\Gamma_{\mu_{1}}\left(  k+q_{4},k+q_{4}+q_{1}\right)  S\left(  k+q_{4}%
+q_{1}\right)  \right.  \label{Box}\\
&&  \left.  \Gamma_{\mu_{2}}\left(  k+q_{4}+q_{1},k-q_{3}\right)  S\left(
k-q_{3}\right)  \Gamma_{\mu_{3}}\left(  k-q_{3},k\right)  \right]  .\nonumber
\end{eqnarray}
The kinematic limit considered by Melnikov and Vainshtein ($q_{1}^{2}\approx
q_{2}^{2}\equiv q^{2}\gg q_{3}^{2})$ is very similar to the case of the pion
transition form factor with highly virtual photons analyzed within N$\chi$QM
in \cite{Anikin:1999cx,Dorokhov:2002iu}. In this limit one has%
\begin{equation}
\mathcal{A}_{\mu_{1}\mu_{2}\mu_{3}\gamma\delta}^{\mathrm{Box}}f^{\gamma\delta
}=-\frac{2\pi^{2}i}{\widehat{q}^{2}}\varepsilon_{\mu_{1}\mu_{2}\delta\rho
}\widehat{q}^{\delta}\epsilon_{4}^{\mu_{4}}\int\frac{d^{4}k}{\pi^{2}}Tr\left[
S\left(  k\right)  \Gamma_{\mu_{4}}\left(  k,k+q_{4}\right)  S\left(
k+q_{4}\right)  \gamma_{\rho}\gamma_{5}S\left(  k-q_{3}\right)  \Gamma
_{\mu_{3}}\left(  k-q_{3},k\right)  \right]  +...,\label{BoxMV}%
\end{equation}
where the higher power corrections are denoted by dots. The result
(\ref{BoxMV}) is proportional to the correlator $T_{\mu_{3}\rho}%
^{a,\mathrm{Loc}}$ which, as discussed above, saturates the full triangle
amplitude at $q_{3}^{2}\gg1$ GeV$^{2}$. Thus, within N$\chi$QM it is the box
diagram which saturates the MV large-${\hat{q}}^{2}$ asymptotics (\ref{MVope})
at $q_{3}^{2}\gg1$ GeV$^{2}\approx\Lambda_{\mathrm{QCD}}^{2}$. At the same
time at $q_{3}^{2}\ll1$ GeV$^{2}$ the box diagram contribution to the
asymptotics is suppressed as it is seen from the behavior of the dashed line
in Fig. \ref{WL}.

In the regime $q_{3}^{2}\ll1GeV^{2}$ the MV asymptotics arises due to the
pion-pole diagram discussed above (Fig. \ref{4by4}b). Indeed, we have%
\begin{eqnarray}
&&\mathcal{A}_{\mu_{1}\mu_{2}\mu_{3}\gamma\delta}^{\pi-\mathrm{Pole}}%
f^{\gamma\delta}   =\left(  \frac{Tr\left[  \widehat{Q}^{2}\lambda
_{a}\right]  }{Tr\left[  \widehat{Q}^{4}\right]  }\right)  ^{2}\frac{G_{P}%
}{1-G_{P}J_{PP}(q_{3}+q_{4})}\\
&&\epsilon_{4}^{\mu_{4}}   \int\frac{d^{4}p}{\left(  2\pi\right)  ^{4}}f\left(
p-q_{3}\right)  f\left(  p+q_{4}\right)  Tr\left[  \Gamma_{\mu_{4}}S\left(
p+q_{4}\right)  i\gamma_{5}S\left(  p-q_{3}\right)  \Gamma_{\mu_{3}}S\left(
p\right)  \right]  \nonumber\\
&&N_{c}   \int\frac{d^{4}k}{\left(  2\pi\right)  ^{4}}f\left(  k-q_{1}\right)
f\left(  k+q_{2}\right)  Tr\left[  i\gamma_{5}S\left(  k-q_{1}\right)
\Gamma_{\mu_{1}}S\left(  k\right)  \Gamma_{\mu_{2}}S\left(  k+q_{2}\right)
\right]  .\nonumber
\end{eqnarray}
The MV limit taken for the second integral corresponds to the asymptotics of the
transition form factor for the pion of fixed virtuality considered earlier in
\cite{Anikin:1999cx,Dorokhov:2002iu}%
\begin{eqnarray}
&&  N_{c}\int\frac{d^{4}k}{\left(  2\pi\right)  ^{4}}f\left(  k-q_{1}\right)
f\left(  k+q_{2}\right)  Tr\left[  i\gamma_{5}S\left(  k-q_{1}\right)
\Gamma_{\mu_{1}}S\left(  k\right)  \Gamma_{\mu_{2}}S\left(  k+q_{2}\right)
\right]  \label{MVlimTri}\\
&&  =\frac{2i}{\widehat{q}^{2}}\varepsilon_{\nu\lambda\delta\sigma}\widehat
{q}^{\delta}N_{c}\int\frac{d^{4}k}{\left(  2\pi\right)  ^{4}}f\left(
k-q_{1}\right)  f\left(  k+q_{2}\right)  Tr\left[  S\left(  k-q_{1}\right)
i\gamma_{5}S\left(  k+q_{2}\right)  \gamma_{\sigma}\gamma_{5}\right]+...
\nonumber\\
&&  =\frac{2i}{\widehat{q}^{2}}\varepsilon_{\nu\lambda\delta\sigma}\widehat
{q}^{\delta}q_{3}^{\sigma}\frac{f_{\pi}\left(  q_{3}\right)  }{Tr\left[
\lambda_{a}^{2}\right]  }+....\nonumber
\end{eqnarray}
Thus the asymptotics of the LbL pion-pole diagram in the MV limit is
\begin{equation}
\mathcal{A}_{\mu_{1}\mu_{2}\mu_{3}\gamma\delta}^{\pi-\mathrm{Pole}}%
f^{\gamma\delta}=\frac{2i}{\widehat{q}^{2}}\varepsilon_{\mu_{1}\mu_{2}%
\delta\rho}\widehat{q}^{\delta}W^{\left(  a\right)  }\Delta T_{\mu_{3}\rho
}+...\label{MVt}%
\end{equation}
where%
\begin{equation}
\Delta T_{\mu_{3}\rho}(q_{3},q_{4})=q_{3}^{\lambda}f_{\pi}\left(  q_{3}%
^{2}\right)  N_{c}\int\frac{d^{4}p}{\left(  2\pi\right)  ^{4}}Tr\left[
\Gamma_{\mu_{3}}S\left(  p+q_{4}\right)  i\gamma_{5}S\left(  p-q_{3}\right)
\Gamma_{\rho}S\left(  p\right)  \right]
  \frac{G_{P}f\left(  p-q_{3}\right)  f\left(p+q_{4}\right)}
{1-G_{P}J_{PP}(q_{3})}.\label{MVtriangle}%
\end{equation}
This expression is similar to $T_{\mu_{3}\rho}^{\mathrm{Pole}}$ in
(\ref{Tpole}). If the renormalization of the axial vertex in (\ref{MVtriangle}%
) coincided with $\Delta\Gamma_{\mu}^{5}(k,k^{\prime})$ in (\ref{GAtot}) then
combining (\ref{BoxMV}) and (\ref{MVt}) one would reproduce the
MV\ asymptotics (\ref{MVlimit}) for all values of $q_{3}^{2}$. However, the
factor $f_{\pi}\left(  q_{3}\right)  $ has a strong dependence on the pion
virtuality $q_{3}^{2}$ and thus the required $1/q_{3}^{2}$ dependence of the
amplitude appears only from the pion-pole factor in (\ref{MVtriangle}) at
$q_{3}^{2}\ll1$~GeV$^{2}$.

The conclusions of this Section are: 1) We exactly reproduce the MV
asymptotics (\ref{MVope}) in the regime $q_{3}^{2}\gg1$~GeV from the quark box
contribution to the LbL amplitude. 2)~In the low-momentum region where effects
of the infrared dynamics may be important we show how this asymptotics arises
from the pion-pole triangle diagram near the position of the pole. 3)~It might
be possible that the MV asymptotics for arbitrary values of $q_{3}^{2}$ is
reproduced within N$\chi$QM if we consider the contact terms contributing to
the four-point amplitude, neglected in the present study. 4)~It follows from
the above results that we partially agree with MV when these authors state that the
\textquotedblleft pion-pole\textquotedblright\ contribution considered in
\cite{Melnikov:2003xd} is an attempt to describe the \textit{complete}, on-
and off-shell light-by-light scattering amplitude in the pseudoscalar
isotriplet channel. 5)~However, the N$\chi$QM model calculations do not support
the factorization ansatz (\ref{PionPole}), with one form factor replaced by a
constant, used by MV in order to simulate this \textit{complete} amplitude%
\footnote{
Moreover, the MV model does not fit the amplitude in other asymptotic
perturbative QCD regimes:
$q_1^2\sim q_2^2\sim q_3^2\gg\Lambda_{\mathrm{QCD}}^2$ and
$q_1^2\sim q_3^2\gg q_2^2\gg\Lambda_{\mathrm{QCD}}^2$. In N$\chi$QM when
the photon virtualities are large
the effective quark propagator connecting the hard photon vertices ($\gamma_\mu$)
becomes the usual massless quark propagator. Thus, the asymptotics of the quark
box diagram (Fig. 3a) is the same as in N$\chi$QM as well as in perturbative
QCD for the massless quark considered in  \cite{Melnikov:2003xd}. In the ENJL model
the quark box diagramm is suppressed at large photon virtualities by additional
VMD form factors.}.
6)~Finally, we agree with \cite{Bijnens:2007pz} that in the MV model
numerically one obtains $a_{\mu,MV}^{\mathrm{\pi^{0},LbL}}=7.97\cdot10^{-10}$
instead of $7.65\cdot10^{-10}$ quoted originally in \cite{Melnikov:2003xd}. Presumably, the discrepancy
has its origin in the numerical treatment of the large-momentum tails in the integrals. With the conformal variables, we treat these tails exactly.

While the inclusion of the box diagram is crucial for the consistency of the
N$\chi$QM, the numerical evaluation of its contribution to LbL is beyond the
scope of this paper. We remark, that in the ENJL model \cite{Bijnens:1995xf}
the result is $2.1(0.3)\cdot10^{-10}$, hence is a few times smaller than the dominant
pion pole term.

\section{Conclusions}

In this paper we have analyzed the leading pion-pole contribution to $a_{\mu
}^{\mathrm{\pi^{0},LbL}}$ in the nonlocal chiral quark model. The basic new
element of our work is the inclusion of the full kinematic dependence of the
pion-photon transition form factors, as it follows from the nonlocal chiral
quark model. The dependence of the form factors on the pion virtuality decreases
the result by about 15\% compared to the case where this dependence is
neglected. We have also demonstrated that the Melnikov-Vainshtein constraints,
necessary for the consistency of the approach with QCD, are satisfied within the model
when the quark box diagram is incorporated. Numerically, we quantitatively confirm
the results obtained in other effective quark models.

An important next step in the investigations of the light-by-light scattering
within nonlocal quark model is to perform an extension to the so-called
\textit{complete} calculation (see
\cite{Bijnens:1995xf,Hayakawa:1997rq,Bartos:2001pg}), whch includes the scalar,
axial-vector, and the $\eta,\eta^{\prime}$ meson exchanges, as well as takes into
account the quark and meson box diagrams. Due to contact terms arising in the
nonlocal model, such a calculation is technically rather involved and will be
presented elsewhere.

\section{Acknowledgments}

We thank S. Eidelman, F. Jegerlehner, N. I. Kochelev, E. A. Kuraev for their
interest in this work. AD thanks for the partial support from Scient. School
grant 195.2008.2 and the JINR Bogoliubov--Infeld program. WB thanks for the
support from the Polish Ministry of Science and Higher Education, grants
N202~034~32/0918 and N~N202~249235.


\end{document}